\documentstyle[aps,pre,multicol]{revtex}
\input{epsf}
\begin{document}
\draft \title{Pulse Bifurcations and Instabilities in an Excitable Medium:
\\
Computations in Finite Ring Domains}
\author{M. Or-Guil$^1$, J. Krishnan$^2$, I. G. Kevrekidis$^2$, M.
B\"ar$^1$\\
\small $^1$Max Planck Institute for the Physics of Complex Systems,\\
\small  N\"othnitzer Stra{\ss}e 38, 01187 Dresden, Germany\\
\small $^2$Department of Chemical Engineering, Princeton University,\\
\small Princeton, NJ 08544, U.S.A.\\ }

\date{\today}

\maketitle

\begin{abstract}

We investigate the instabilities and bifurcations of traveling
pulses in a model excitable medium;
in particular we
discuss  three different scenarios for the loss of stability {\it resp.}
the disappearance of
stable pulses.
In numerical simulations beyond the instabilities we
observe replication of pulses
(``backfiring'') resulting in complex periodic or spatiotemporally
chaotic dynamics as well as modulated traveling pulses.
We  approximate the linear stability of
traveling pulses through computations in a
finite albeit  large  domain with periodic
boundary conditions.
The critical eigenmodes at the onset of the instabilities
are related to the resulting spatiotemporal dynamics and 
``act'' upon the back of the pulses.
The first scenario has been analyzed earlier [M. G. Zimmermann
{\it et al.}, Physica D {\bf 110}, 92 (1997)]
for high excitability {\it resp.} low excitation threshold:
it involves the collision of a
stable pulse branch with an unstable pulse branch in a so called
T-point.
In the frame of traveling wave ordinary
differential equations, pulses correspond
to homoclinic orbits
and the T-point to a double heteroclinic loop.
We investigate this transition for a pulse in a domain with 
finite length and periodic boundary conditions.
Numerical evidence of the proximity of the infinite-domain T-point
in this setup appears in the form of two saddle-node bifurcations.
Alternatively, for intermediate excitation threshold,
an entire cascade of saddle-nodes causing a ``spiraling'' of
the pulse branch appears near the parameter values
corresponding to the infinite domain T-point.
Backfiring appears at the first saddle-node bifurcation, which limits
the existence region of stable pulses. 
The third case found in the  model for large excitation threshold 
 is an oscillatory instability giving rise to ``breathing'', traveling pulses
which periodically vary in width and speed.

\end{abstract}

\pacs{PACS numbers:
03.40.Kf, 
47.54.+r, 
82.40.Ck  
}

\begin{multicols}{2}

\section{Introduction}

One-dimensional excitable media exhibit nonlinear traveling waves such
as wave trains and solitary pulses.
Examples include concentration waves in chemical reactions
in solution and on surfaces \cite{RossScience,KapSho,CO-exp},
signal propagation in neurons and in cardiac tissue \cite{keener98}.
These processes are often modelled by reaction-diffusion (RD) equations
of activator-inhibitor type \cite{keener98,Mikhailov}.
A pulse is a localized structure; it may result from a finite amplitude
perturbation of a linearly stable rest state.
The pulse shape usually decays exponentially as a function of the
distance from the pulse center.
Pulses can be analytically approximated in the asymptotic limit where
the dynamics of the activator
is much faster than that of the inhibitor variable, and where the activator
variable diffuses, but the inhibitor does not.
There, it can be shown that a  pulse
exists and is stable \cite{Mikhailov}.

More recently, spatiotemporal chaos has been found in a variety
of excitable model systems.
Several examples have been reported in models whose kinetics
possess three distinct homogeneous
steady states (fixed points); apart from the stable rest state, these
media also exhibit two unstable fixed points
\cite{baer94,merkin96,argentina97}.
One example arises in a model for catalytic CO oxidation,
where the inhibitor is a so-called surface structure variable; its
kinetic nullcline displays
a sigmoidal shape - threshold dependence on the activator variable (adsorbate
coverage) \cite{CO-model}.
In contrast to the standard linear dependence of the inhibitor nullcline on
the activator, this functional form leads to the additional fixed points
mentioned above.
Similar behavior is often seen in models in physiology describing
the dynamics of the membrane voltage (activator) controlled by so
called gating variables for the ion channels (inhibitor).
The kinetics of these gating variables display again a threshold-type,
sigmoidal dependence on
the membrane voltage, often leading to three (spatially homogeneous)
fixed points for the PDE \cite{keener98}.
If only a single gating variable is involved, these models are qualitatively
similar to the equations for catalytic CO oxidation studied here;
a good example
is provided by the Morris-Lecar model used to describe the membrane
potential in a barnacle muscle fiber  \cite{morris81}.

If the inhibitor kinetics are fast enough, the CO oxidation model displays
an instability that has been colloquially named backfiring of pulses
\cite{baer94}.
Related phenomena include the wave-induced chemical chaos found in
the Gray-Scott model for an autocatalytic chemical reaction \cite{merkin96},
as well as complex behavior in amplitude equations describing dynamics near
a
Takens-Bogdanov (TB) point \cite{argentina97}.
In addition, even reaction-diffusion media whose kinetics are characterized
by a single fixed point, sometimes display pulse instabilities and
backfiring
under excitable conditions \cite{mimura97,christoph}.
An analysis of the traveling wave ordinary differential equations derived
from the original PDEs for the CO oxidation
model reveals the basic mechanism for the destruction of stable pulses
leading to complex behavior: it involves a  so-called T-point \cite{Swedes}
as well as a rich web
of bifurcations of pulse solutions \cite{Krishnan}.
More recently, a similar bifurcation structure has been found in a
model for intracellular calcium waves in pancreatic acinar
cells \cite{Sneyd00} and in the ultrarefractory version of the
FitzHugh-Nagumo model \cite{Romeo01}.
The resulting complex dynamics is often governed by a coherent structure
described as a ``wave emitting front'' \cite{Swedes}, a nonlinear front
involving a spatially uniform {\it unstable} state that invades
a spatially uniform stable one.
The unstable state behind the propagating front evolves
into spatiotemporal chaos.
Similar behavior has been seen in the amplitude equations in the
neighborhood of the TB point,
and was named  ``chaotic nucleation'' by those authors
\cite{argentina97}.

In this paper we investigate instabilities and bifurcations
in a model of a catalytic surface reaction  \cite{CO-model}.
Using a computer-assisted approach we calculate solution branches
 in large, finite domains, under periodic boundary conditions
(which we will, with slight abuse of terminology,  refer to as pulses) and
perform linear stability analysis of these pulse solutions.
The eigenvalues of the corresponding Jacobian  reflect the  growth exponents
of perturbation modes.
The emphasis of the present paper is on the form of the spectra and
in particular on the shape of the critical modes;
this is in contrast to earlier work which focused
mostly on the pulse bifurcation structure \cite{Swedes,Krishnan}.

A homogeneous steady state of reaction-diffusion equations
in an infinite system possesses a continuous spectrum,
reflecting the growth rates of perturbation modes.
The eigenmodes are harmonic functions, whose wavenumber
parametrizes the corresponding eigenvalues.
In a very long or infinite system, a single pulse can
conceivably be considered (in the appropriate norm) as a
perturbation of this uniform solution.
Under these circumstances, the modes far away from the
pulse are still delocalized, harmonic wave modes; the
continuous spectrum remains essentially unchanged.
Furthermore, additional
perturbation modes exist that are localized at the site of the pulse and
decay exponentially away from it.
The corresponding eigenvalues are discrete, in contrast
to the continuous band of eigenvalues belonging to the nonlocalized modes.
In infinite or periodic systems,
the Goldstone mode is an example of such a localized mode.
In a spatially homogeneous system it is given  by the
spatial derivative of the pulse profile, and accounts
for a shift in space (due to translational invariance);
the corresponding eigenvalue is  zero.

A bifurcation or instability of a given solution
upon change of a single control parameter
is typically accompanied by
a single real eigenvalue or a pair of complex conjugate eigenvalues crossing
the
imaginary axis.
For a real eigenvalue, a saddle-node bifurcation is the most common
case, while alternative bifurcations like the transcritical or
pitchfork bifurcations require certain symmetries for the equations and
the solution.
The case of a drift
pitchfork bifurcation of a stationary pulse
was studied in  \cite{kness92}.
A destabilizing perturbation grows in an oscillatory manner if the
eigenvalues
are complex and in a non-oscillatory manner if the corresponding eigenvalue
is real.
The first case leads to a Hopf bifurcation, while the latter case may
correspond
either to a saddle-node or to a transcritical or pitchfork bifurcation.

\begin{figure}
\epsfxsize=80mm
\centerline{\epsffile{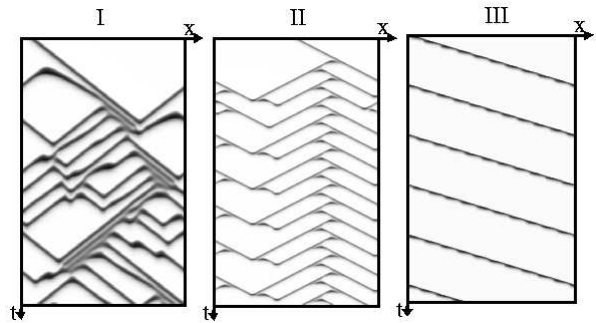}}
\vspace{2mm}
\caption{ Space-time plots from numerical integration of Eqs. (1)
showing the time evolution of pulses at parameter values beyond
the instability onset. A stable pulse solution for a subcritical
parameter value, characterized by rest state $A$, was used as the
initial condition.
I: Backfiring in the immediate vicinity of a T-point.
The resulting behavior is non-periodic for the given system length
and initial conditions.
II: Backfiring after a saddle-node bifurcation.
The resulting behavior is
periodic in time for the given system length and initial conditions.
III: A supercritical Hopf bifurcation leads to modulated traveling
waves. The pulse shows a periodic  oscillation of shape and
speed; variations  appear mainly at its trailing edge.
Black denotes high values of $u$, white corresponds to $u = 0$.
Parameters:
I: $b = 0.07$, $\epsilon = 0.1075$,$L = 100$,  $\Delta T = 119.2$,
II: $b = 0.15$, $\epsilon = 0.0931$, $L = 100$, $\Delta T = 238.42$,
III: $b = 0.2$, $\epsilon = 0.062$, , $L = 50$, total integration time
$\Delta T = 238.42$. }
\label{fig1}
\end{figure}

Three examples of dynamics where localized initial
conditions beyond the uniform solution excitation threshold
do {\it not} eventually evolve into stable
pulses, are shown in Fig.~\ref{fig1}.
For high excitability {\it resp.} small excitation threshold,
pulse-like initial conditions
evolve in a traveling pulse that splits off new pulse-like
excitations traveling in the opposite direction.
This is the phenomenon termed backfiring in \cite{baer94}.
It can eventually evolve in spatiotemporally chaotic (I)
or complex periodic (II) fashion.
For large excitation threshold, modulated traveling pulses are also seen
(III).
Their shape undergoes a breathing, periodic variation.

\section{Model and Methods}

We investigate a model of activator-inhibitor type originally derived
for CO oxidation on Pt(110) \cite{CO-model},  in a
parameter regime where the kinetics give rise to one stable and two unstable
steady homogeneous solutions.
This model is related to the FitzHugh-Nagumo system
and describes the interaction
of a fast activator $u$ and a slow inhibitor variable $v$:
\begin{eqnarray} \label{eqn}
\partial_t u&=& \partial_x^2 u + {1\over\epsilon} u (1-u)
(u-\frac{b+v}{a})  \nonumber\,,\\
\partial_t v &=& f(u)-v \label{pde} \,,\\ f(u) &=&
\left\{\begin{array}{ll} 0,& 0 \le u <1/3\\
            1-6.75u(u-1)^2,& 1/3\le u \le 1\,,\\ 1,& 1 < u
\end{array} \right. \nonumber
\end{eqnarray}
with $x \in [0,L]$ and periodic boundary conditions.

The time scales ratio $\epsilon > 0$ is used as the control parameter.
The case $\epsilon \rightarrow 0$ corresponds to the aforementioned
asymptotic limit where stable pulses are expected.
The parameter $b$ controls the
excitability threshold of the system: the value of $b$ is proportional to
the
magnitude of the critical perturbation  which will trigger a pulse.
The value of $a$ is fixed at $0.84$ throughout the paper.
We will vary $b$, thus varying the excitability.
In the parameter range considered here, three relevant fixed points exist:
the stable state  $A=(0,0)$, the saddle $B=(b/a,0)$, and unstable
focus $C$.

We  shall now examine the stability of pulses traveling in a background of
the stable rest state $A$.
Since we investigate solutions moving with fixed velocity $c$,
the analysis of their stability is performed in the comoving frame
$z:=x-ct$:
\begin{eqnarray}\label{CoMov}
\partial_t u & = & \partial_z^2 u + c \partial_z u +
\frac{1}{\epsilon}u(1-u)(u-\frac{b+v}{a}) \,,\\
\partial_t v & = &   c \partial_z v + f(u) - v \, .\nonumber
\end{eqnarray}
In this frame, traveling waves with speed $c$ correspond  to
time independent, steady solutions.
Linearization of  these equations around a stationary solution
$u_0(z), v_0(z)$ yields an eigenvalue problem  for
small perturbations  $(r(z,t),s(z,t)) \propto (r(z),
s(z))e^{\lambda t}$:
\begin{eqnarray}
{\cal M} \left(\begin{array}{l} r(z) \\ s(z) \end{array} \right)
= \lambda \left(\begin{array}{l} r(z) \\ s(z) \end{array} \right)
\nonumber \,,\\ \nonumber \\
{\cal M} = \left(\begin{array}{cc} \partial_z^2 +
c \partial_z + g_1(z) & g_2(z)\\
\partial_u f\;(u_0) & c \partial_z - 1 \end{array}\right)
\label{linear}
\end{eqnarray}
with
\begin{eqnarray}
g_1(z) &=& -\frac{1}{\epsilon} \left(u_0 (u_0 -1) + (u_0 -
\frac{b+v_0}{a})(2
u_0 -1) \right) \,,\nonumber\\
g_2(z) &=& \frac{u_0}{\epsilon a} (u_0-1).
\nonumber
\end{eqnarray}
The linear stability problem amounts to the determination of the
spectrum of the Jacobian $\cal M$ in (\ref{linear}).
For the homogeneous steady states $A$ or $B$, at least one of the
off-diagonal matrix elements is zero. Thus, the diagonal elements of the
matrix $\cal M$
suffice to compute the spectrum. For the $A$ steady state it is:

\begin{equation}\label{EVA}
\lambda_{A,1} = -\frac{b}{\epsilon a} - k^2 + i c k, \quad \lambda_{A,2} =
-1+ i c k,\\
\end{equation}
and for the $B$ steady state it is:
\begin{equation}\label{EVB}
\lambda_{B,1} =   -\frac{b}{\epsilon a} \left( \frac{b}{a} - 1  \right)  -
k^2 + i c k, \quad \lambda_{B,2} = -1+ i c k,\\
\end{equation}
where $k$ is the wavenumber of the perturbation. In the case of periodic
boundary conditions studied here,
$k = n \frac{2\pi}{L}$ applies.
The real part of the eigenvalues $\lambda_2$ is $-1$.
The eigenvectors are then $(r, s)^T = (1,0)^T e^{ikz}$ and
$(0,1)^T e^{ikz}$.

In general, traveling wave solutions $u_0(z), v_0(z)$
and the eigenfunctions of the Jacobian $\cal M$ are not available
in closed form. Thus, the problem has to be approximated numerically.
We approximate pulses by computing steady
solutions in a finite length system and 
the traveling frame through a pseudospectral discretization of
Eqs.\ (\ref{CoMov}) with periodic boundary conditions
and Newton-Raphson iterations.
The velocity $c$, which is not known {\it a priori}, is formally
an additional variable along with  the Fourier coefficients
of the solution.
One additional pinning condition  singles out one of the
infinitely many solutions existing due to translational invariance and
allows the numerical computation of the speed.
The eigenfunctions and the spectrum  of the Jacobian
are obtained in Fourier space
resulting from a 200 ({\it resp.} 250 for case I)
mode decomposition of the stationary solution.
The zero eigenvalue, which always exists
due to the translational symmetry of the problem,
is used as a numerical accuracy check and has been obtained with a precision
of $10^{-3}$ or better.
The time evolution of unstable solutions was computed using an
explicit finite difference scheme to solve Eqs.\ (\ref{eqn}), discretizing space in 1024 points
and using a time step of $\Delta t = 0.0122$.


Because traveling  solutions fulfill the conditions $\partial_t u =
\partial_t v=0$
in the comoving frame, they can also be obtained in the
traveling wave ordinary differential equations (TWODE) following
from Eqs.\ (\ref{CoMov}):
\begin{eqnarray} \label{ODE}
\frac{du}{dz} &=& w \,,\nonumber\\
\frac{dw}{dz} &=& -c w + {1\over\epsilon} u (u-1)
(u-u_{th})\,,\label{twode}\\
\frac{dv}{dz} &=& (v-f(u))/c \,,\nonumber
\end{eqnarray}
with $u_{th}$ = $(b+v)/a$.
In this frame, a homogeneous solution corresponds to a fixed point,
a pulse  to a homoclinic orbit, and a front to a heteroclinic orbit.
Consequently, in the parameter range studied here, three relevant fixed
points exist:
${\cal A}=(0,0,0)$, ${\cal B} =(b/a,0,0)$, and the focus $C$.
\begin{figure} 
\epsfxsize=55mm
\centerline{\epsffile{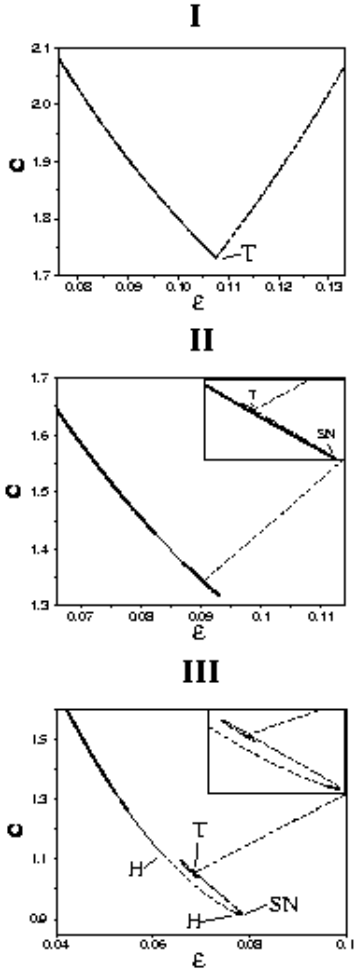}}
\vspace{2mm}
\caption{ Pulse speed as a function of the parameter $\epsilon$. I:
  $b=0.07$, II: $b=0.15$, and III: $b=0.2$.
Thick lines are pulses with rest state $A$ (i.e. pulses in a large box
with periodic boundaries representative of pulses to the state $A$ on
the infinite line corresponding to the homoclinic orbits to ${\cal
A}$); 
thin lines are pulses with rest state $B$.
Solid (dashed) lines denote stable (unstable) branches.
A T marks our approximation of the double heteroclinic connection
point (the T-point) where pulses with rest state $A$ ``collide'' with
pulses with rest state $B$.
H denotes Hopf bifurcations, SN denotes saddle-node
bifurcations.
In the cases II and III, the branch of pulses with rest state $A$
spirals into the T-point.
This is not the case for I, compare Fig.\ \ref{fig5}b.}
\label{fig2}
\end{figure}

\section{Results}

We consider three cases, at increasing values of the excitation
threshold, controlled through the parameter $b$ at fixed $a=0.84$:
Case I ($b$ = 0.07), case II ($b$= 0.15) and case III ($b$ = 0.20).
We proceed as follows:
first, branches of pulse solutions on a ring, representative of ,,true"
pulse solutions in a infinite domain, 
are presented for the relevant range of the
control parameter $\epsilon$; we characterize these pulses by their
speed $c$ (Fig.~\ref{fig2}, \ref{fig5}b).
We then show pulse profiles and spectra at selected values of $\epsilon$
shortly before and after the onset of instability
(Figs.\ \ref{fig5}, \ref{fig4}, \ref{fig3}),
and present the destabilizing mode $(r, s)^T$
(Figs.\ \ref{fig8}, \ref{fig4}, \ref{fig3}).
Representative post-instability spatiotemporal dynamics
can be found in Fig.\ \ref{fig1}.

Fig.~\ref{fig2} shows pulse speed as a function of the parameter $\epsilon$
for the three cases.
The thick lines correspond to stable pulses with background state $A$,
while dashed lines correspond to unstable
pulses with background state $B$.
The transition point between the two families is the so called
T-point \cite{Glendinning}, denoted by a T.
This point is defined as a double heteroclinic connection
between the fixed points $A$ and $B$ in the frame of the traveling
wave ODEs, Eqs.\ (\ref{ODE}).
In the frame of the original equation (\ref{eqn}),
these heteroclinic orbits correspond to fronts.
The branch of pulses corresponding to homoclinic orbits to ${\cal A}$
in the TWODE may (Fig.\
\ref{fig2}.II, III) or may not (Fig.\  \ref{fig2}.I)
spiral into the T-point.
Spiraling is observed when two of the eigenvalues of the
linearization of the TWODE around the fixed point ${\cal B}$ are
complex conjugate \cite{Swedes}.

\subsection{Case I}     

Fig.~\ref{fig2}.I shows speed as a function of
the parameter $\epsilon$  for $b = 0.07$ computed for pulses 
in a large system with periodic boundaries.
The picture appears at first glance identical with the result
found in a continuation of homoclinic orbits in the TWODE \cite{Swedes}. 
The branch 
of pulses to  $A$ (thick line) does not spiral into the T-point.
To be exact, the T-point only exists for
pulses on the infinite line; 
what -- at the resolution of our picture -- still appears as
a T-point will be discussed in more detail below.
To understand the nature of the instability in this case,
we focus on the solutions near the T-point
value $\epsilon_T \approx 0.10744$.

As mentioned above, for an infinite system, the T-point corresponds to
a double heteroclinic connection in the traveling wave ODEs, Eqs.\
(\ref{ODE}).
Close to $\epsilon_T$ the orbits homoclinic to ${\cal A}$
also approach the fixed point ${\cal B}$ (and vice versa);
this approach to ${\cal B}$ makes the dynamics
in $z$ progressively slower.
Thus, close to $\epsilon_T$, pulses with rest state $A$ (resp. $B$) 
will also exhibit extensive regions close to $B$ (resp. $A$),
practically giving rise to {\it fronts} between $A$ and $B$
within the pulse profile.
When we perform a continuation {\it for a pulse in a ring of finite length},
the space on the ring is apart from the excitation plateau
divided between residence close to $A$ and
residence close to $B$,  so that the total period is constant as we vary
$\epsilon$. 

\begin{figure}  
\epsfxsize=80mm
\vspace{2mm}
\centerline{\epsffile{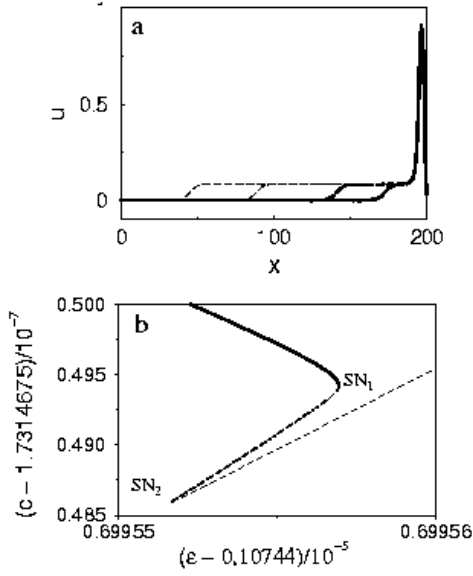}}
\vspace{2mm}
\caption{
a)  Four pulse solutions on the ring from the middle branch
in (b) illustrating
the gradual transformation from solutions identifiable with
pulses with rest state $A$ ($u=0$) to
unstable solutions identifiable with pulses with rest state $B$ ($u=0.083$).
The  $B$ domain in the waveform becomes wider upon decrease of
$\epsilon$, along the middle branch in (b).
b) Bifurcation diagram, with respect to $\epsilon$, of pulses 
 for case I, exhibiting two saddle-nodes.
The thick solid line can be  associated with
stable pulses homoclinic to $A$; the thin dashed line can be associated with unstable pulses homoclinic
to $B$.
The thick dashed line corresponds to the transition region between
these two cases; it constitutes the incarnation (for our finite, large
ringlength continuation) of the T-point.
When the upper saddle-node bifurcation ($SN_1$) takes place, our finite
ringlength solution could be still described as an approximation of
a pulse to $A$ but with a small $B$ shoulder; the
converse description holds at the lower saddle-node bifurcation ($SN_2$).
It is reasonable to consider as most representative of the
infinite-domain T-point the location, along this thick-dashed line,
where the pulse solution contains comparable large patches 
close to $A$ and close to $B$ -- 
roughly the middle of this middle-branch.
}
\label{fig5}
\end{figure}
To study this behavior, numerical continuation techniques are needed.
We computed the stationary solutions of Eqs.\ (\ref{CoMov}) on a ring
of length $L = 200$ with 250 modes, 1024 collocation points and a
parameter step size $\Delta \epsilon = {\cal O}(10^{-10})$.
Periodic approximations of pulse solutions with rest state $A$ are shown
in Fig.\ \ref{fig5}a
as they approach the T-point and undergo the gradual transition to
 pulse solutions with rest state $B$.
One can clearly recognize the increasing domain  $B=(b/a, 0)$ 
at the back of the pulse.
As the state $B$ is unstable, one might expect that the 
pulse in the ring  loses stability as soon as the $B$ domain gets large enough.
However, one should be aware of the fact that the $B$ domain is
moving with the speed of the pulse.
It is possible that the $B$ domain is only convectively unstable
in the comoving frame.
In other words, the perturbations growing on the $B$ state in a
stationary frame may spread slower than the pulse motion,
and the pulse in an infinite domain will be stable.
A mathematically precise description of this phenomenon has been
given by Sandstede and Scheel \cite{SS00}.
A related result  has been obtained by Nii \cite{Nii00}, who shows that
eigenvalues accumulate in the area bounded by the essential spectra
of $A$ and $B$.
The opposite case, {\it i.e.} perturbations on the $B$ plateau spread
faster
than the pulse speed, is described in the next subsection (case II).

\begin{figure}  
\epsfxsize=80mm
  \centerline{\epsffile{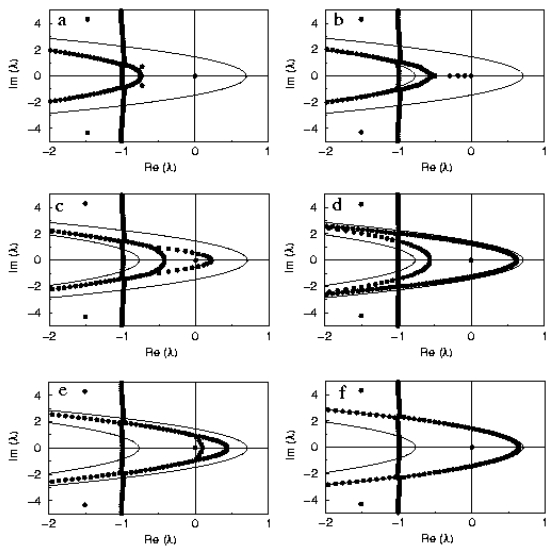}}
\vspace{2mm}
  \caption{Spectra of pulses in case I for different values of $\epsilon$.
  The solid lines correspond to the continuous spectrum of the rest state
$A$,
  Eqs.\ \ref{EVA}
  (left parabola) resp.\ rest state $B$ (right parabola), Eqs.\ \ref{EVB}.
  The
  pictures show the typical change of the spectrum along the solution branch
c-$\epsilon$.
  Parameters: $a=0.84$, $b=0.07$, $L=200$ ($L=400$ for picture d),
  a) $\epsilon = 0.106425824$,
  b) $\epsilon = 0.107446965$,
  c) $\epsilon = 0.107446995547$,
  d) $\epsilon = 0.1079$,
  e) $\epsilon = 0.107446995527$,
  f) $\epsilon = 0.109936154$.
  }
  \label{fig6}
  \end{figure}

Typical spectra  are shown in Fig.\ \ref{fig6}
for solutions along the $c-\epsilon$-branch.
To facilitate a better comparison,
the continuous spectra of the rest states $A$ (Eqs.\ (\ref{EVA})) and
$B$ (Eqs.\ (\ref{EVB})) are depicted as solid lines;
the computed eigenvalues are denoted by circles.
Fig.\ \ref{fig6}a  shows the spectrum of a solution
approximating a stable pulse with rest state $A$, while
Fig.\ \ref{fig6}f shows that of a solution approximating
an unstable pulse with rest state $B$.
In both cases, the parameter $\epsilon$ is ``far''
from the T-point bifurcation and
the eigenvalues  belonging to nonlocalized eigenvectors
compare  well with the
continuous spectrum of the respective rest state.
Figs.\ \ref{fig6}b-e show the gradual transition between the two cases.
Fig.\ \ref{fig6}b shows the spectrum for a stable solution for which
the domain $B$ is just starting to appear in the back of the pulse.
Its width is $L_B \approx  5$.
Several discrete  eigenvalues have appeared on the negative real axis;
a similar spectrum has been
predicted and found  by Sandstede and Scheel \cite{SS00}.

The details of the spectrum tranformation are as follows:
Initially (far to the left of the T-point in Fig. \ref{fig2}a), 
the pulse spectrum contains two discrete complex conjugate eigenvalues
(which remain more or less unchanged through the bifurcation diagram) 
as well as two ``pieces" of continuous spectrum: a parabola, and a vertical line.
For our calculations in the finite domain, initially the
parabola {\it does} have a 
``tip"
(an eigenvalue on the real axis) while the vertical line {\it does not} have
an eigenvalue on the real axis.
As we vary $\epsilon$ the parabola approaches the straight line; 
eventually the parabola appears as {\it crossing} the line. 
While this occurs, a number of complex conjugate pairs of eigenvalues
emerge from the parabola and line approximating the essential
spectrum.
Fig. \ref{fig6}a shows the situation where the first of these pairs
has appeared. 
Upon further increase of $\epsilon$, these pairs collapse on
the negative real axis where they split and become real. 
These real eigenvalues lie in  the {\it
absolute} spectrum of state $B$ (see \cite{SS00}); 
they appear when the solution has a visible $B$ ``shoulder" behind the
pulse, see Fig. \ref{fig6}b.  
As $\epsilon$ varies these real eigenvalues
become complex conjugate.
Several complex eigenvalue pairs of the parabola 
give the appearance of a secondary parabola  - 
this is the first intimation of what will become the essential spectrum of $B$.
Only two of the ``momentarily real" eigenvalues remain real 
- one of them forms the tip of the ``new parabola", while the 
other forms the tip of the ``old parabola".

Gradually, as the $B$ plateau grows, two distinct phases of spectrum
movement are observed.
First this new $B$-associated parabola moves to the right in the
complex plane and at some point it starts crossing the imaginary axis.
That is precisely the first saddle-node bifurcation we observe - the critical eigenvalue is the tip of this secondary parabola, whose origin we just discussed, see Fig.\ \ref{fig6}c.
Continuing further on the middle, already unstable branch,
the $B$-domain gets wider and many secondary (apparently Hopf)
bifurcations occur as the
eigenvalues of this ``discretized" parabola cross successively the
imaginary axis.
When both $A$- and $B$-plateaus are equally present, one expects to
see echoes of both $A$- and $B$-continuous spectra in the solution
spectrum, and that is indeed seen in
Fig.\ \ref{fig6}d.
Near the lower saddle-node bifurcation, the $B$-plateau is almost as
wide as the system length.
There, one may describe the solution more reasonably as an
approximation to a pulse with rest
state $B$ with a short $A$ domain at its front.
The eigenvalue spectrum at this situation is depicted in Fig.\
\ref{fig6}e.
The second (lower) turning point occurs when the ``discretized" parabola
of eigenvalues that corresponded initially to continuous spectrum of $A$
moves to the right and  the real eigenvalue at the tip
of this discretized parabola crosses the imaginary axis.

Eventually, the old $A$-parabola eigenvalues merge with the
$B$-parabola as the $A$-shoulder gradually disappears.
Fig.\ \ref{fig6}f shows the spectrum for a solution approximating a pulse
with rest state $B$ far from the T-point bifurcation for a high values
of $\epsilon$.
In the end, the two ``tips" (the old and the new one) collide and
become again a complex conjugate pair - the final pure-$B$ parabola has no real tip.

Sandstede and Scheel in \cite{SS00} have proven that the spectrum of
pulses  with rest state  $A$ with
a ``shoulder" of $B$ state as in Fig.\ \ref{fig5}a is comprised
of the essential spectrum of the homogeneous background state $A$ plus
eigenvalues lying in the absolute spectrum \cite{AbSp}
of $B$ located on the real axis and bounded by a  maximum value which is
negative.
We deviate from their picture, 
when the $B$ shoulder becomes relatively large. 
In other words, their result on the infinite line 
is reproduced by our numerical
stability analysis in the ring, if the condition $L_B << L$ holds.
The spectrum transformations, however, 
reflect the influence of the periodic boundary conditions and the
finite length. 
Because of the violation of the above assumption near the T-point
(see Fig. 3a) the spectra we find
for a large box are not well described as the
union of the essential spectrum of $A$ and the absolute
spectrum of $B$.
The results of \cite{SS00} also imply that if one considers
 pulses on the infinite line instead of the large
wavelength approximation with periodic boundary conditions 
used here, one will find a monotonic
approach of the pulse branches to the T-point.
In their scenario no eigenvalues are crossing the real axis upon approaching
the T-point (in contrast to the computations on the ring) and
no bifurcation (like the saddle-node bifurcation found for the
ring) can occur.

Focusing on the first instability ($SN_1$),
the critical eigenmode corresponding to the saddle-node appears
localized at the back of the pulse (see Fig. 5).
In addition, it is acting mainly on the activator (Fig. 5b); we observe that new pulses can now split off the existing pulse (``backfiring").
Fig.\ \ref{fig8} shows the time evolution beyond the saddle-node
bifurcation in a comoving frame.
The initial condition is the stable pulse just before the
bifurcation.
Fig.\ \ref{fig8}a shows how a perturbation with support over the
$B$-plateau
grows with time, while the shape of the preceding front essentially does not
change.
The back (what starts as the  $B$-plateau)  grows until its maximum
reaches the excited state $u=1$.
The excited  domain widens, but its plateau is unstable
against oscillating perturbations \cite{Swedes}.
Therefore, the situation rapidly evolves, and the growing
excited domain breaks down, giving rise to two new seeds for
$A$-pulse like entities.
This behavior has been described by the term
``backfiring'' \cite{baer94,Swedes}.
Backfiring occurs repeatedly, and the newly generated
pulse-like structures annihilate upon collision with similar
objects traveling in the opposite direction;
this interplay of instability, new pulse generation and annihilation
gives rise
to the non-periodic space-time behavior shown in Fig.\ \ref{fig1}.I.
\begin{figure}  
 \epsfxsize=80mm
  \centerline{\epsffile{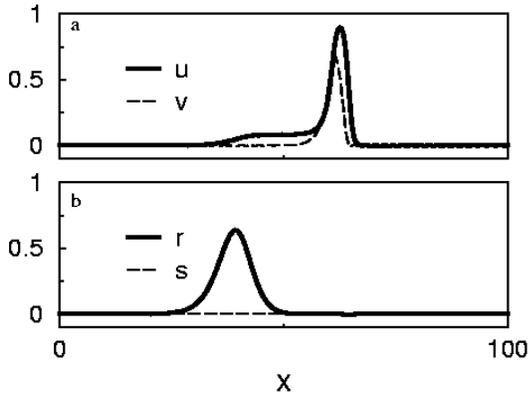}}
\vspace{2mm}
\caption{ a) Unstable pulse solution in case I in the vicinity of the
saddle-node bifurcation $SN_1$, Fig.\ \ref{fig5}b.
b) Destabilizing eigenmode. Parameters: $a=0.84$, $b=0.07$,
$\epsilon=0.1074469956$, $L=200$. Only part of the domain is shown.
}
  \label{fig7}
  \end{figure}

\begin{figure}[t]  
 \epsfxsize=80mm
  \centerline{\epsffile{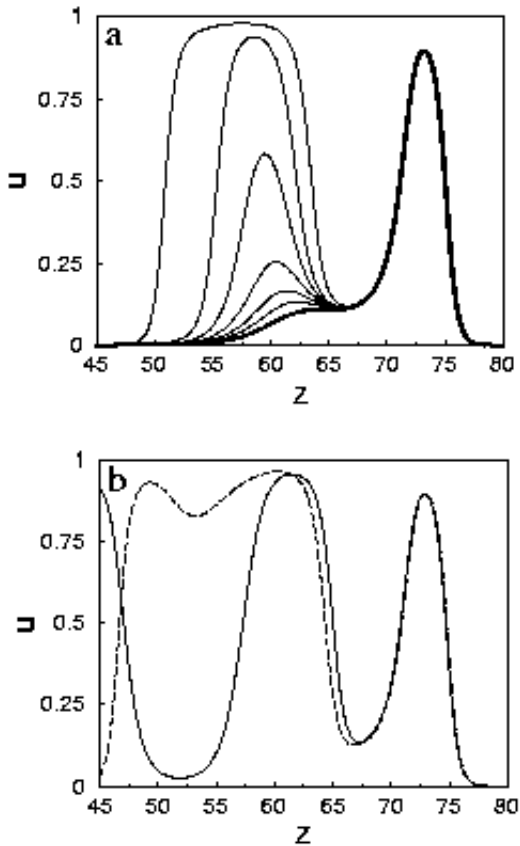}}
 \vspace{2mm}
  \caption{Time evolution  for a value of $\epsilon$
  after the saddle-node bifurcation $SN_1$ in a comoving frame.
The initial condition is the stable pulse-like solution just
before the bifurcation
(thick black line in a)).
The thin lines in a) demonstrate how the shape  of the solution
changes only at the back of the pulse-like solution.
A localized perturbation grows in amplitude and width in the course of
time and approaches the rest state $C$, where $u=1$.
Further evolution of the dynamics is given by the  dashed line
in picture b).
Due to oscillatory instability of the homogeneous state $C$, the
corresponding domain cannot become large.
A breakdown leads to the generation of two further pulse-like states
traveling in opposite directions (thin solid line).
  }
  \label{fig8}
  \end{figure}

Altogether, our results give an interesting and well resolved
picture of stability of pulses on a large finite
ring near the T-point.
In contrast, earlier computations \cite{Swedes,Krishnan} have
used coarser steps in the continuation algorithms and smaller domains.
The present computations show how the 
pulse solutions change
in a gradual fashion;  and that two distinct solutions branches show an
extremely narrow hysteresis; the middle branch mediates
the transition between the two different pulse types.
The main changes appear in an extremely tiny region of
parameter space and can therefore only be captured with 
careful numerics.
The spectrum transformations near parameters that exhibit 
a T-point in the corresponding TWODE  appear in a similar 
fashion for case II and III studied below. 
There, however, the T-point is not involved in the disappearance 
of stable traveling pulses through instability and bifurcation. 
Thus, we will not discuss it in these cases. 

\subsection{Case II}     %

\begin{figure}  
\epsfxsize=80mm
\centerline{\epsffile{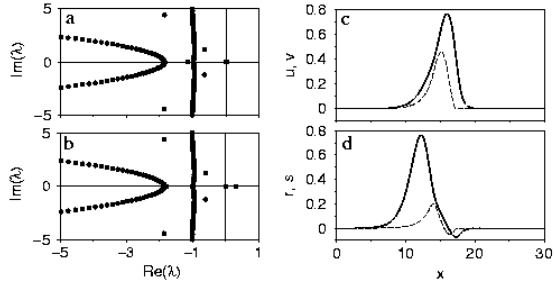}}
\vspace{2mm}
  \caption{
  Case II: a) Eigenvalues at $\epsilon = 0.0927$ before a saddle-node
bifurcation of a pulse decaying into the rest state $A$;
  b) eigenvalues at $\epsilon = 0.09299$ after this bifurcation. A discrete
eigenvalue crosses the imaginary
  axis. c) Unstable  solution and
  d) corresponding destabilizing eigenmode $(r, S)^T$ at $\epsilon =
0.09299$. Parameter: $a=0.84$, $b = 0.15$, $L=100$.
  }
  \label{fig4}
  \end{figure}

Fig.\ \ref{fig2}.II shows the $c-\epsilon$ diagram for $b=0.15$.
As in case I, we again observe here the transformation from
pulses with rest state $A$ to
 pulses with rest state $B$ at what appears like a T-point.
The branch corresponding to ``pulses to $A$"
spirals into this T-point.
This behavior is caused by imaginary eigenvalues of the fixed point
$B$ in the TWODE at the T-point conditions, and has been predicted
from
general arguments \cite{Glendinning}.
Thus, the branch of initially stable pulses with rest state $A$
undergoes a sequence of saddle-node bifurcations upon approaching the
T-point.
This sequence of saddle-node bifurcation is consistent with the
results of Sandstede and Scheel \cite{SS00}.
The spectrum of $A$-pulses near the T-point in the infinite system
approaches the union of the essential spectrum of $A$ and the absolute
spectrum of $B$.
In this case, however, the absolute spectrum of $B$ contains real 
{\it positive} eigenvalues.
One may, therefore expect infinitely many saddle-node bifurcations 
as the T-point is approached.
In the stability-relevant first saddle-node bifurcation,
the solution branch of stable wavetrains with rest state $A$
turns around and becomes unstable.
As is required for a saddle-node bifurcation,
a single real eigenvalue crosses the imaginary axis.
This can be seen in Fig.\ \ref{fig4}, left column.
The two spectra correspond to pulse-like solutions
before and after the saddle-node bifurcation along the solution branch.
The unstable pulse  and  the destabilizing eigenmode
are shown in Fig.\ \ref{fig4}, right  column.
Once more, the destabilizing mode affects primarily the
back of the pulse.
It is worth noting, that upon further continuation the branch of
pulses to $A$ spirals towards the T-point related situation over 
a cascade of saddle-node bifurcations. 
Each saddle-node bifurcations adds an additional positive eigenvalue 
to the spectrum. 
This is in contrast to the study in \cite{hcchang}, where a single 
eigenvalue repeatedly crosses zero along a spiraling pulse branch
with a cascade of saddle-node bifurcation.  

Numerical simulations of the model Eqs.\ \ref{eqn} shortly after the
saddle-node bifurcation, for values of $\epsilon$ for which
no solution with rest state $A$ exists, exhibits the
phenomenon we termed above ``backfiring''.
In the transients, our pulse-like object generates near its
back other pulse-like entities  traveling in the opposite direction,
see Fig.\ \ref{fig1}.II.
After this transient period, the resulting spatiotemporal pattern
in our finite domain becomes periodic in time.
Simulations show, though, that this observation depends on the initial
conditions: non-periodic patterns like the one shown in Fig.\ \ref{fig1}.I
may also appear for the same parameter values.

\subsection{Case III}   

\begin{figure}  
\epsfxsize=80mm
\centerline{\epsffile{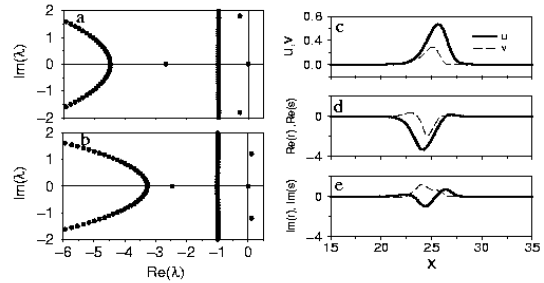}}
\vspace{2mm}
\caption{ Case III: a) Eigenvalues at $\epsilon = 0.053$ before a Hopf
bifurcation;
  b) eigenvalues at $\epsilon = 0.072$ after this bifurcation. A discrete
pair of complex conjugated
  eigenvalues crosses the imaginary
  axis. c) Unstable pulse-like solution as well as d) real and e)
imaginary
part of the corresponding
  destabilizing eigenmode
  $(r, s)^T$ at $\epsilon = 0.065$. Parameters: $a=0.84$, $b = 0.2$, $L=50$}
  \label{fig3}
  \end{figure}

Fig.\ \ref{fig2}.III shows the $c-\epsilon$ 0 for $b=0.2$.
Here, the pulse solution  with rest state $A$ becomes unstable
through a Hopf bifurcation even before the first saddle-node
bifurcation is reached. 
The eigenvalue spectrum is shown in Fig.~\ref{fig3}, left column,
on both sides of this Hopf bifurcation.
It can be seen that one discrete pair of complex
conjugate eigenvalues crosses the imaginary axis.
The second column of Fig.~\ref{fig3} shows the pulse-like
solution {\it after} the bifurcation
as well as the real and imaginary parts of the critical eigenmode
$(r, s)^T$.
Note that the perturbation due to the eigenmode is localized and
has its main contribution -- once more -- at the back of the pulse.
As the corresponding
eigenvalues are complex conjugate, the resulting perturbation
oscillates with time
at a frequency given by the imaginary part of the eigenvalues.

Numerical integration of the model Eqs.\ (\ref{eqn}) shows that the Hopf
bifurcation which
leads to destabilization is supercritical \cite{Krishnan}.
The resulting pattern 
after destabilization consists of a {\it modulated} traveling pulse which
oscillates
in time, especially at its  back (compare the simulation shown in  Fig.\
\ref{fig1}.III.)


\section{Conclusion}

We have investigated the transition from stable pulse propagation to
various regimes of more complicated spatiotemporal dynamics, namely
modulated pulses, periodic and chaotic pulse backfiring.
In all three cases, the transition can be explained by either a Hopf
instability (modulated pulses) or a saddle-node bifurcation
(leading to backfiring) of the stable pulse solution.
In all cases, the transition is connected with either a single or a pair of
complex conjugate discrete eigenvalue(s) with zero real part(s).
In a finite domain with periodic boundary conditions --
the typical experimental setup for investigation of pulses --
spectra change continuously near the T-point 
in the fashion described in case I.
The form of the corresponding critical
eigenmode(s) allows some insight into how pulses become
unstable {\it resp.} evolve in space and time.
The dynamics in general still contains mostly propagating localized
pulse-like structures whose evolution is governed by the unstable
eigenmode(s)
in the Hopf case (modulated) or by the critical eigenmode of the
saddle-node bifurcation.
Typically, the critical eigenmodes have support at the back of the pulse.

The results here should carry over to models with similar phenomenology
mentioned in the introduction.
Preliminary results \cite{zimmermann} show that the transition to
wave-induced chemical chaos in the Gray-Scott model \cite{merkin96}
is also caused by a saddle-node bifurcation of pulses near a T-point.
T-points can only be found in systems with multiple
homogeneous fixed points ({\it e.g.} one stable rest state and two
additional
unstable steady states).
The complex behavior seen in media with a single stable fixed point
\cite{mimura97,christoph} may be caused through a different mechanism.
For a model of the catalytic NO-CO reaction, upon change of 
control parameter first modulated
traveling waves are seen and then periodic backfiring is found.
This cannot be due to a T-point, but may instead be caused by a
global bifurcation of the periodic modulated pulses - a scenario
already suggested in a study of the present model with different
control parameters \cite{Krishnan}.
Finally, it is important to note that a simple instability of a
finite wavelength pulse solution,
like the Hopf bifurcation in case III, only leads to a modulation of
the shape, while a saddle-node bifurcation limits the existence of
a certain type of pulses and may give rise to replication of pulse-like
structures (cases I, II). The role of saddle-node bifurcations in
the replication of pulses and creation of space-time defects
has recently been investigated in the Gray-Scott model
\cite{Nishiura99} and in the Complex-Ginzburg-Landau equation
\cite{Brusch00}.

{\bf Acknowledgement:} We thank Bj\"orn Sandstede and Arnd Scheel
for enlightening discussion regarding Ref. \cite{SS00}. IGK
acknowledges the support of the NSF (USA) and of the A.v.Humboldt
Stiftung (Germany).

\end{multicols}

\begin{thebibliography}{99}
\bibitem{AbSp}
The absolute spectrum is defined in \cite{SS00}. For the example
investigated here, the absolute spectrum of $B$ at the T-point lies
on the real line and is given by the interval $[-0.77564,-0.03847]$.
%
For other parameters, the absolute spectra lies on the interval
$[-\frac{b}{a\epsilon}, -\frac{c^2}{4} + \frac{b}{a\epsilon}
( 1 - \frac{b}{a} )]$.
\end{thebibliography}
\end{document}